\documentstyle[12pt,a4,epsf]{article}
\newcommand{\BC}{\begin{center}}    \newcommand{\EC}{\end{center}}
\newcommand{\BE}{\begin{equation}}  \newcommand{\EE}{\end{equation}}
\newcommand{\BA}{\begin{eqnarray}} \newcommand{\EA}{\end{eqnarray}}
\newcommand{\BN}{\begin{enumerate}} \newcommand{\EN}{\end{enumerate}}
\newcommand{\BI}{\begin{itemize}}   \newcommand{\EI}{\end{itemize}}
\newcommand{\BFig}{\begin{figure}}  \newcommand{\EFig}{\end{figure}}
\newcommand{\BTab}{\begin{table}}   \newcommand{\ETab}{\end{table}}
\newcommand{\BT}{\begin{tabular}}   \newcommand{\ET}{\end{tabular}}
\newcommand{\BM}{\begin{minipage}}  \newcommand{\EM}{\end{minipage}}

\newcommand{\LL}{\label}
\newcommand{\NN}{\nonumber}

\newcommand{\VS}{\rule[-2mm]{0mm}}   % use \FB{\VS{Xmm} text}
\newcommand{\HS}{\hspace*}

\newcommand{\LRA}{\Longrightarrow}

\def\JL#1#2#3#4{{#1} {\bf #2} (#3) #4}   % new
\newcommand{\HL}{\hline}                 % new
\newcommand{\MB}{\mbox}                  % new
\begin{document}
%%%%%%%%%%%%%%%%%%%%%%%%%%%%%%%%%%%%%%%%%%%%%%%%%%%%%%%%%%%%%%%%%%%%%
\begin{flushright}
PSI-PR-96-35 \\
October 1996 
\end{flushright} 
\VS{10mm}
\BC
{\LARGE Annihilation into Channels with Strangeness and \\
  the OZI Rule Violation}%
\footnote{Invited contribution at LEAP-96,
          August 1996, Dinkelsb\"uhl, Germany \\
\HS{8mm}(to be published in Nuclear Physics B Proceedings Supplement).}

\VS{10mm}
{\sc V.E.~Markushin}
\\[5mm]
{\it Paul Scherrer Institute, 5232 Villigen PSI, Switzerland}
\EC

\VS{10mm}
%%%%%%%%%%%%%%%%%%%%%%%%%%%%%%%%%%%%%%%%%%%%%%%%%%%%%%%%%%%%%%%%%%%%%%%%
\begin{abstract}
Two-step mechanisms in the $N\bar{N}$ annihilation and their role
in the OZI rule violating reactions are discussed. 
In particular the two meson rescattering mechanism for $\pi\phi$ channel
including all off-shell effects is typically two orders of magnitude bigger
than the OZI tree level expectation and explains the observed ratio $\phi
\pi/\omega \pi$ in the annihilation at rest.
The rates for the final states including photons, $\gamma\omega$ and
$\gamma\phi$, can be explained in the vector dominance model. The observed
rate for $p\bar{p}\to\gamma\omega$ is suppressed due to destructive
interference between the intermediate  $\rho$ and $\omega$ states while the
interference in $p\bar{p}\to\gamma\phi$ is required to be constructive
leading to a large ratio $\gamma\phi/\gamma\omega$.
\end{abstract}
%%%%%%%%%%%%%%%%%%%%%%%%%%%%%%%%%%%%%%%%%%%%%%%%%%%%%%%%%%%%%%%%%%%%%%%%%%

%%%%%%%%%%%%%%%%%%%%%%%%%%%%%%%%%%%%%%%%%%%%%%%%%%%%%%%%%%%%%%%%%%%%%%%%%
\section{Introduction}
\label{Intr}

In the nucleon-antinucleon annihilation near threshold, the channels with
strangeness contribute only a few percent \cite{CCSppbar}, nevertheless, the
corresponding reactions are of big importance because they provide  new
information on reaction mechanisms, proton structure, and exotic states. Of
special interest are the processes resulting in the violation of the
Okubo-Zweig-Iizuka (OZI) rule \cite{Ok63,Zw64,Ii66}.  The recent
experiments at LEAR have been very successful in providing the data on this
subject
\cite{AST91,AST93,CBC93a,CBC93b,CBC93c,CBC94,CBC95,CBC95a,OBEL1,OBEL2,JETSET,JETSET95,Wi96,Lu96}. 
A characteristic feature of the low energy $p\bar{p}$ annihilation is
an abundant production of the $\phi$ mesons in some channels
($\phi\pi$, $\phi\gamma$, $\phi\rho$, $\phi\pi\pi$, $\phi\omega$, $\phi\phi$)
which is expected to be OZI suppresses on the tree level because
the $\phi$ is dominantly a $\bar{s}s$ configuration.  

  Because the OZI rule can be dynamically broken, it is important to
understand whether the effects observed in the $p\bar{p}$ annihilation can
result from known mechanisms of the OZI violation.
% before firm conclusions about `new physics' can be made. 
% There are different mechanisms of the OZI rule breaking, such as, two-step
% processes with ordinary hadrons, which are essential due the unitarity,
% \cite{Li84} and the production of flavor mixed or exotic (glueballs, hybrids)
% states.
% Different mechanisms of the OZI rule breaking can be considered.
%
% The production of flavor mixed or exotic (glueballs, hybrids) intermediate 
% states may also be considered. 

Two-step processes with ordinary hadrons as intermediate doorway states 
have been considered long ago and were shown to be important already in
unitarity approximation (see \cite{Li84} and references therein).  In the
case of $N\bar{N}$ annihilation two meson doorway mechanisms have been
studied for various final states containing $\phi$ mesons in $p\bar{p}$ and
$p\bar{d}$ reactions which violate the OZI rule
\cite{LZL,LLZ,LL,BL94,BL94a,BL95,BL95a,GLMR}.
Another way of violating the OZI rule is production of flavor mixed or
exotic (glueballs, hybrids) intermediate states, an approach which will
not be discussed here.  
Furthermore, one can link the OZI rule violation to the nucleon structure 
by introducing an intrinsic $\bar{s}s$ component as in \cite{EKKS} (see also
\cite{GYF} and references therein for further details). 

%The two-step processes are essential due the unitarity, and the way the OZI
%rule avoids large unitarity corrections in some cases is the long-studied
%problem of its own \cite{L2,GI,LZ}. 

The goal of this paper is to review recent theoretical results concerning
the evaluation of conventional two step mechanisms for $p\bar{p}\to\phi X$
as an explanation of the OZI rule violation.
`New physics' should be introduced only if these explanations fail. 
For earlier reviews we refer to \cite{Lo95,Sa95}.

%%%%%%%%%%%%%%%%%%%%%%%%%%%%%%%%%%%%%%%%%%%%%%%%%%%%%%%%%%%%%%%%%%%%
\section{The OZI Rule and Flavor Mixing}

The OZI rule \cite{Ok63,Zw64,Ii66} states that processes with disconnected
quark lines in initial or final states are suppressed. For the heavy quarks
($c,b,t$) this suppression is a consequence of the asymptotic freedom of QCD
because the coupling constant $\alpha_s(q^2)$ becomes small, therefore the
creation of a quark-antiquark pair is a perturbative process. The case of
strange quarks belongs to the non-perturbative domain and involves
more complicated dynamics.

One reason for the OZI rule violation is the flavor mixing in ordinary
(nonexotic) hadrons. In QCD with massless $u,d,s$ quarks the hadrons would
form $SU(3)$ multiplets, with the OZI suppression mechanism being trivially
avoided for the states which are a superposition of all possible
configurations. In reality the $s$ quark is significantly heavier than the
$u$ and $d$ quarks, and the $SU(3)$ symmetry is broken.
In particular, the physical states $\omega$ and $\phi$ are the
superpositions of the $SU(3)$ singlet $\omega_1$ and octet $\omega_8$
states \cite{Ok63,AFO}:
\BA
  \left( {\phi \atop \omega} \right) & = &
  \left( {\cos \Theta \atop \sin \Theta} \ \ {-\sin \Theta \atop \cos \Theta}
  \right) \left( {\omega_8 \atop \omega_1} \right)
  \\
    \omega_1 & = & \frac{u\bar{u} + d\bar{d} + s\bar{s}}{\sqrt{3}}
               =  \sqrt{\frac{2}{3}}q\bar{q} + \sqrt{\frac{1}{3}}s\bar{s}
  \\
    \omega_8 & = & \sqrt{\frac{1}{3}}q\bar{q} - \sqrt{\frac{2}{3}}s\bar{s}
\EA
where $q\bar{q} = (u\bar{u} + d\bar{d})/\sqrt{2}$. 
The so called ideal mixing corresponds to a complete decoupling of the
$\bar{s}s$ and $\bar{q}q$ components:
\BA
  \Theta_i = \arctan \sqrt{1/2} = 35.3^{\circ} \LRA \ \
  {\phi = s\bar{s} \atop \omega = q\bar{q} }
\EA

%%%%%%%%%%%%%%%%
\VS{-20mm}
\BC \mbox{\epsfysize=50mm\epsffile{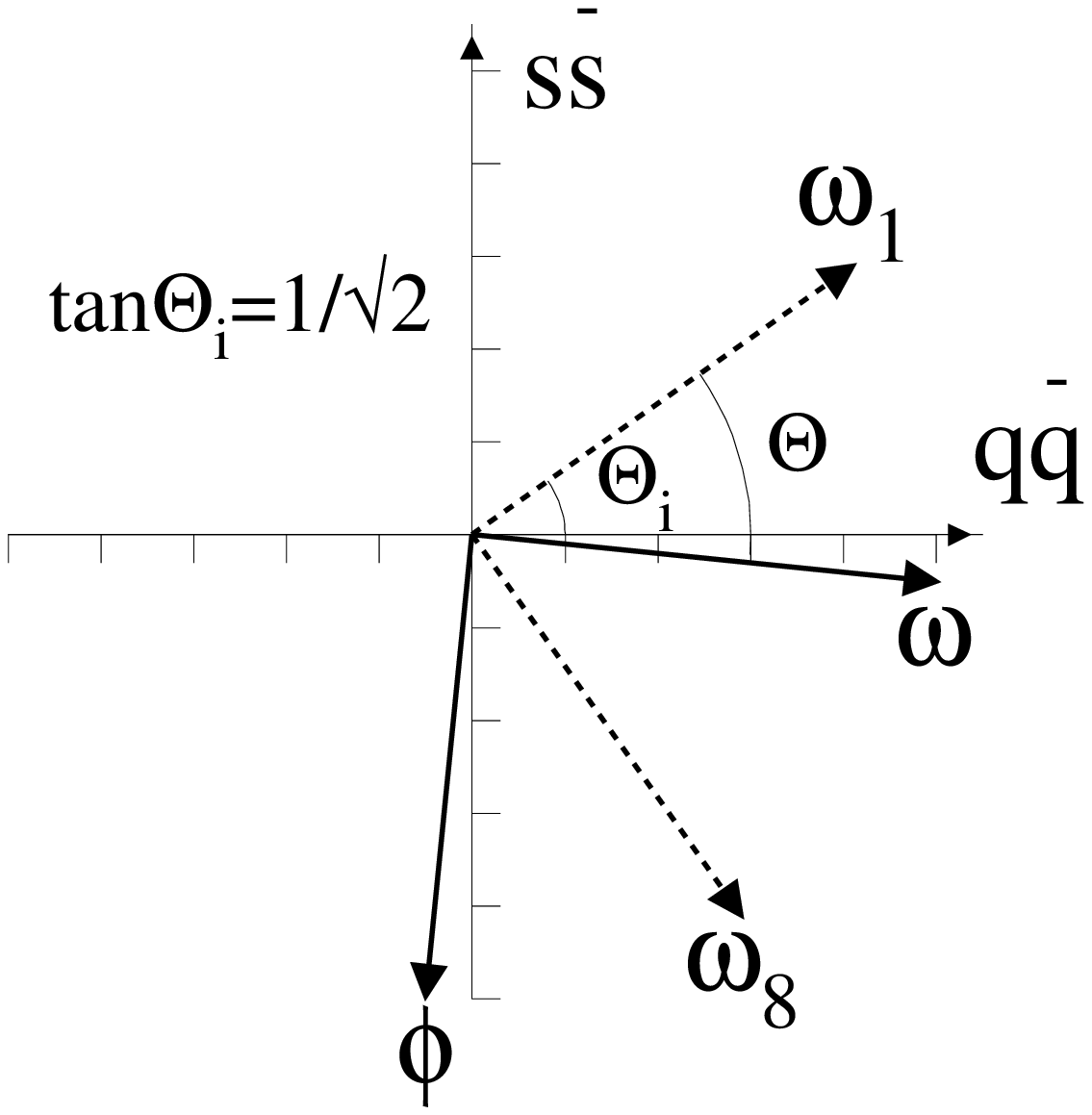}} \EC
\VS{-20mm}
%%%%%%%%%%%%%%%%

The Gell-Mann--Okubo mass formula with octet-singlet mixing 
%\BA
%  M_{88}^2 = \frac{1}{3}(4 m_{K^*}^2 - m_{\rho}^2) (1+\Delta)
%\EA
gives the mixing angle $\Theta \approx 39^{\circ}$ \cite{PDT}. 
The mixing angle can also be obtained from the partial widths
of the $\omega$ and $\phi$ decays into $e^+e^-$
\BA
  \frac{\Gamma_{\omega\to e^+e^-}}{M_{\omega}} :
  \frac{\Gamma_{\phi\to e^+e^-}}{M_{\phi}} = \tan^2 \Theta
\EA
leading to $\Theta = 37.1^{\circ}$.

Since $\Theta \not = \Theta_i$  the OZI forbidden processes involving the
$\phi$ production  can go via the $s\bar{s}-q\bar{q}$ mixing. The
corresponding amplitudes are expected to be proportional to the deviation
from the ideal mixing $(\Theta - \Theta_i)^2 \sim (1 - 4.2) \cdot 10^{-3}$
for $\Theta - \Theta_i \approx 1.8^{\circ} - 3.7^{\circ}$. The OZI rule
violation on the level of the flavor mixing is usually called {\em
nondramatic}.
Compared to the level given by the $\omega-\phi$ mixing, the low energy
$p\bar{p}$ annihilation into the $\gamma\phi$, $\pi\phi$, and $\phi\phi$
channels appears to be rather {\it dramatic\/} exceeding it by 1-2 orders of
magnitude. This problem will be discussed in detail in the following
sections.

%%%%%%%%%%%%%%%%%%%%%%%%%%%%%%%%%%%%%%%%%%%%%%%%%%%%%%%%%%%%%%%%%%%%%%%%%%
\section{The OZI Rule Violation in the $p\bar{p}$ Annihilation
         into Two Vector Mesons}

Table~\ref{TabBRVV} shows the summary of the experimental data  on the
$p\bar{p}\to 1^- + 1^-$ reactions at rest (the annihilation in liquid target
predominantly occurs in the $S$-wave, in gaseous targets both the $S$ and
$P$ waves contribute; the LX trigger in gas mainly corresponds to the
$P$-wave annihilation).

% Check the P <-> gas/LX ???
%%%%%%%%%%%%%%%%%%%%%%%%%%%%%%%%%%%%%%%%%%%%%%%%%%%%%%%%%%%%%%%%%%%%%%%
\BTab
\caption{\label{TabBRVV} The experimental branching ratios for $p\bar{p}$
annihilation at rest into $\rho\omega$, $\omega\omega$, $\gamma\omega$,
$\rho\phi$, $\omega\phi$, $\gamma\phi$, channels.}
\BC
\HS{-2mm}\BT{|c|c|c|l|}
\HL
 Reaction &    BR   &   Condition  &  Ref. \\
\HL
$p\bar{p}\to\rho\omega$ &  $5.4(6) \cdot 10^{-2}$ & gas. & %ASTERIX 
                                                           \cite{AST93} \\
%                       &  $5.8(9) \cdot 10^{-2}$ & gas. & \cite{AST93} 
                        &  $3.0(7) \cdot 10^{-2}$ &  S   & %ASTERIX 
                                                           \cite{AST93} \\
                        &  $6.4(11)\cdot 10^{-2}$ &  P   & %ASTERIX 
                                                           \cite{AST93} \\
%                       &  $1.9(4) \cdot 10^{-2}$ &  S?  & \cite{Weid90} \\
%                       &  $6.4(13)\cdot 10^{-2}$ &  P?  & \cite{Weid90} \\
                        &  $2.3(2) \cdot 10^{-2}$ & liq. & %Bizzari et al. 
                                                           \cite{Biz69} \\
\HL
$p\bar{p}\to\omega\omega$& $3.32(34) \cdot 10^{-2}$ & liq. & %CBC 
                                                           \cite{CBC93b}\\
                        &  $1.4(6) \cdot 10^{-2}$ & liq. & %Bloch et al. 
                                                           \cite{BFL70} \\
\HL
$p\bar{p}\to\gamma\omega$& $6.8(18)\cdot 10^{-5}$ & liq. & %CBC 
                                                           \cite{CBC93a} \\
\HL\HL
$p\bar{p}\to\rho\phi$   &  $3.4(8) \cdot 10^{-4}$ & gas. & %ASTERIX 
                                                           \cite{AST91} \\
                        &  $4.4(12)\cdot 10^{-4}$ & gas./LX & %ASTERIX 
                                                              \cite{AST91} \\
                        &  $3.4(10)\cdot 10^{-4}$ & $^1S_0$ & %ASTERIX 
                                                              \cite{AST91} \\
                        &  $3.7(9)\cdot 10^{-4}$  & $^3P_J$ & %ASTERIX 
                                                              \cite{AST91} \\
\HL
$p\bar{p}\to\omega\phi$ &  $6.3(23)\cdot 10^{-4}$ & liq.    & %Bizzari et al. 
                                                              \cite{Biz71} \\
                        &  $3.0(11)\cdot 10^{-4}$ & gas.    & %ASTERIX 
                                                              \cite{AST91} \\
                        &  $4.2(14)\cdot 10^{-4}$ & gas./LX & %ASTERIX 
                                                              \cite{AST91} \\
                        &  $5.3(22)\cdot 10^{-4}$ & $^1S_0$ & %ASTERIX 
                                                              \cite{AST91} \\
                        &  $2.9(14)\cdot 10^{-4}$ & $^3P_J$ & %ASTERIX 
                                                              \cite{AST91} \\
\HL
$p\bar{p}\to\gamma\phi$ &  $1.7(4) \cdot 10^{-5}$ & liq. & %CBC 
                                                           \cite{CBC95} \\
\HL
\ET
\\
%a) $3.4(8) \cdot 10^{-4} = 1.6(5)\cdot 10^{-4}(S) + 1.8(6)\cdot 10^{-4}(P)$
%b) $4.4(12)\cdot 10^{-4} = 0.8(3)\cdot 10^{-4}(S) + 3.6(11)\cdot 10^{-4}(P)$
\EC
\ETab
%%%%%%%%%%%%%%%%%%%%%%%%%%%%%%%%%%%%%%%%%%%%%%%%%%%%%%%%%%%%%

The vector dominance model (VDM) can be used to connect the amplitude of the
reaction $p\bar{p}\to\gamma X$ with the amplitudes for
$p\bar{p}\to\rho X$ and  $p\bar{p}\to\omega X$ \cite{VDM,LLZ,CBC93a,CBC93b}. 
The final states of interest for the OZI rule violation
correspond to $X=\phi,\omega$.

%%%%%%%%%%%%%%%%%%%%%%%%%%%%%%%%%%%%%%%%%%%%%%%%%%%%%%%
\BFig
\VS{-30mm}
\BC \mbox{\epsfysize=40mm \epsffile{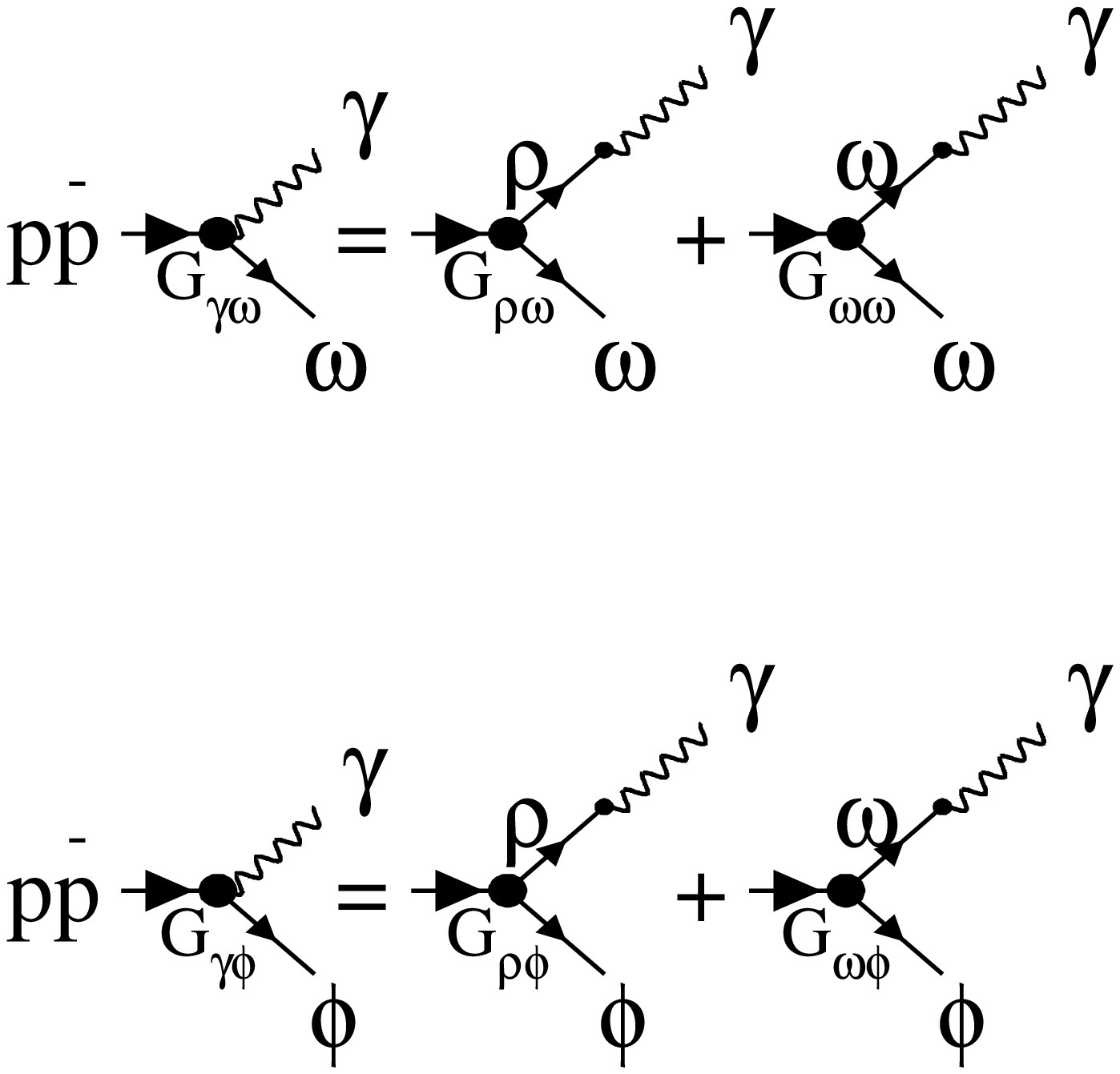}} \EC
\VS{-50mm}
\caption{\LL{FigVDM} The amplitudes of the reactions
$p\bar{p}\to\gamma\omega$ and $p\bar{p}\to\gamma\phi$ in the
vector dominance model.} 
\VS{-30mm}
\EFig
%%%%%%%%%%%%%%%%%%%%%%%%%%%%%%%%%%%%%%%%%%%%%%%%%%%%%%%

The VDM relation has the form  (see Fig.~\ref{FigVDM}) 
\BA
  G_{\gamma\omega} & = & 
  \frac{\gamma_{\rho\gamma}}{m_{\rho}^2}
  G_{\rho\omega} F_{\rho\omega} + 
  \frac{\gamma_{\omega\gamma}}{m_{\omega}^2}
  G_{\omega\omega} F_{\omega\omega} 
\label{VDM1}
\\
    G_{\gamma\phi} & = & 
  \frac{\gamma_{\rho\gamma}}{m_{\rho}^2}
  G_{\rho\phi} F_{\rho\phi} + 
  \frac{\gamma_{\omega\gamma}}{m_{\omega}^2}
  G_{\omega\phi} F_{\omega\phi} 
\label{VDM2}
\EA
where the amplitudes $G_{ab}$ correspond to the final states $ab$ and 
the coupling constants $\gamma_{\gamma\rho}$ and $\gamma_{\gamma\omega}$
are related by 
\BA
    \gamma_{\gamma\rho} = 3 \gamma_{\omega\rho} = 
    m_{\rho}^2 \frac{e}{f_{\rho}} = 0.054 m_{\rho}^2 \ .
\label{VDMg}
\EA
The form-factors $F_{ab}(p_a^2)$ describe the off-shell
behavior of the \MB{$p\bar{p}\to ab$} vertex,  
they are normalized by the condition
\mbox{$F_{\omega\omega}(m_{\omega}^2)=F_{\rho\omega}(m_{\rho}^2)=
F_{\omega\phi}(m_{\omega}^2) = F_{\rho\phi}(m_{\rho}^2)=1$}. 
  
  Since the final states have well defined $C$-parity $C=+1$, the
allowed initial $p\bar{p}$ states are $^1S_0$ and $^3P_J$. 
Our analysis will be focused on the $S$-wave annihilation, therefore
only the $^1S_0$ state contributes.
The corresponding partial width is
\BA
    \Gamma_{ab} & = & \frac{G_{ab}^2 P_{ab}^3}{4 \pi}
\label{Gamma}  
\EA
where $P_{ab}$ is the CMS momentum of the particles $a$ and $b$ in
the final state. 

From (\ref{VDM1}--\ref{Gamma}) and neglecting
the mass difference of the $\omega$ and $\rho$ mesons the following result
for the branching ratios is derived \cite{VDM} 
\BA
   & \frac{BR(\gamma\omega)}{BR(\rho\omega)} = \HS{60mm} & \\ & =
    \left|  F_{\rho\omega} + \frac{\sqrt{2}}{3} 
      \sqrt{\frac{BR(\omega\omega)}{BR(\rho\omega)}}
      F_{\omega\omega} e^{i \beta_{\omega}} \right|^2
      \left( \frac{e}{f_{\rho}} \right)^2
      \left( \frac{P_{\gamma\omega}}{P_{\rho\omega}} \right)^3  
   & \NN \\
   & \frac{BR(\gamma\phi)}{BR(\rho\phi)} = \HS{60mm} & \\ & = 
    \left| F_{\rho\phi} + \frac{1}{3} 
      \sqrt{\frac{BR(\omega\phi)}{BR(\rho\phi)}}
      F_{\omega\phi} e^{i \beta_{\phi}} \right|^2
      \left( \frac{e}{f_{\rho}} \right)^2 
      \left( \frac{P_{\gamma\phi}}{P_{\rho\phi}} \right)^3  
   & \NN
\EA
where $\beta_X$ is the relative phase between the two terms
corresponding to $\rho X$ and $\omega X$ intermediate states
($X=\omega,\phi$).    
 
%%%%%%%%%%%%%%%%%%%%%%%%%%%%%%%%%%%%%%%%%%%%%%%%%%%%%%%%%%%%%%%%%%%%%%%
\subsection{Reaction $p\bar{p}\to\gamma\omega$}

Using the experimental branching ratios 
$BR(\gamma\omega)/BR(\rho\omega) = (2.3 \pm 1.1)\cdot 10^{-3}$  and 
$BR(\omega\omega)/BR(\rho\omega) = 1.1 \pm 0.3$
\cite{AST93,CBC93a,CBC93b}  
one gets
\BA
    \left|  F_{\rho\omega} + (0.49 \pm 0.07)  
      F_{\omega\omega} e^{i \beta_{\omega}} \right| = 0.49 \pm 0.13
\LL{DRgammaomega}
\EA

If $F_{\rho\omega} \approx F_{\omega\omega} \approx 1$,
then $\beta_{\omega} \approx \pi$, i.e. 
there is a {\em destructive} interference between the $\rho$ and $\omega$
intermediate states with isospin $I=1,0$ in the total amplitude of
the reaction $p\bar{p}\to\gamma\omega$. 
The result of the analysis by the Crystal Barrel Collaboration \cite{CBC93a}
obtained with a model formfactor \cite{FF} is
$\cos \beta_{\omega} = -0.60 \pm 0.18$.

%%%%%%%%%%%%%%%%%%%%%%%%%%%%%%%%%%%%%%%%%%%%%%%%%%%%%%%%%%%%%%%%%%%%%%%
\subsection{Reaction $p\bar{p}\to\gamma\phi$}

Using the experimental branching ratios 
$BR(\gamma\phi)/BR(\rho\phi) =  0.05 \pm 0.025$ and \\
$BR(\omega\phi)/BR(\rho\phi) =  1.6 \pm 1.1$ 
\cite{AST91,CBC95} 
one gets
\BA
    \left|  F_{\rho\phi} + (0.4 \pm 0.1)  
      F_{\omega\phi} e^{i \beta_{\phi}} \right| & = & 1.1 \pm 0.3
\LL{DRgammaphi}
\EA
If $F_{\rho\phi} \approx F_{\omega\phi} \approx 1$, then
the data can be explained by a {\em constructive} interference between
the $\rho$ and $\omega$ terms.    
The effect of the formfactor in this case can be more significant than
for the $\gamma\omega$ final state because of a larger difference
in the final momenta: 
$P_{\rho\phi}=0.28\;$GeV/c and $P_{\gamma\phi}=0.66\;$GeV/c
(compare with the case $p\bar{p}\to\rho\omega,\gamma\omega\ $:
$P_{\rho\omega}=0.53\;$GeV/c and $P_{\gamma\omega}=0.78\;$GeV/c). 

The analysis done in \cite{CBC93b}  arrived at the conclusion that the VDM
prediction is too small to describe the experimental data. This conclusion,
however, strongly depends on the model formfactor \cite{FF}.
If we use the following formfactor 
\BA
   F_{\rho\phi}(p_{\rho}^2) =
   \frac{\Lambda^2+P_{\rho\phi}^2}{\Lambda^2+P^2(p_{\rho}^2)}
\EA
where $P(p_{\rho}^2)$ is the CMS momentum of the particles in the final
state, and a similar one for the $\omega\phi$ vertex, 
than for $\Lambda=1\;$GeV/c the suppression factor is 
$F_{\rho\phi}(0)=0.75$ and the VDM relation (\ref{DRgammaphi})
holds true for $\cos\beta_{\phi}\approx 1$.  

Thus the relative phases of the intermediate terms $I=0,1$ for the two
reactions $p\bar{p}\to \gamma\omega, \ \gamma\phi$ are different, and this
agrees with the observation that  $\omega - \phi$ mixing cannot be the
dominant mechanism of the $\phi$ production.
Indeed, the OZI rule violation in the $\phi\omega$ and $\phi\rho$ channels,
given by the ratios  $BR(\phi\omega)/BR(\omega\omega) \sim 
BR(\phi\rho)/BR(\omega\rho) \sim 10^{-2}$, seems to exceed the level
expected from the deviation from the ideal mixing,\footnote{The estimation
can vary if the final state phase space factor is taken into account.}
$(\Theta-\Theta_i)^2 \sim
0.4\cdot 10^{-2}$, but is not dramatic. Thus a seemingly drastic violation
of the OZI rule observed in  the ratio
$BR(\gamma\phi)/BR(\gamma\omega) = 0.24 \pm 0.09$ \cite{CBC93a,CBC95,CBC95a} 
results from partial cancellation of the two terms $I=0,1$ for the
$\gamma\omega$ channel and the constructive interference between the
intermediate states for the $\gamma\phi$ channel.
Therefore the core problem lies in the explanation of the
$p\bar{p}\to\rho\phi$ and $p\bar{p}\to\omega\phi$ reactions which will be
discussed in Sec.~\ref{TSrhophi}. 
 
%The following picture summarizes the results for the S-wave
%annihilation into two vector mesons.\\ 
%\FB{$p\bar{p} \to 1^- + 1^-$}
%\BC
%\FB{\PB{55mm}{ $\gamma\omega$ \\ $6.8(18)\cdot 10^{-5}$}}
%\\[5mm]
%\FB{\PB{55mm}{ $\omega\omega$ \\ $3.3(3)\cdot 10^{-2}$}}
%\HS{5mm} VDM \HS{5mm} 
%\FB{\PB{55mm}{ $\rho\omega$   \\ $\sim 5\cdot 10^{-2}$}}   % gas/P
%% \FB{\PB{55mm}{ $\rho\omega$   \\ $1.9(4)\cdot 10^{-2}$}} % liq.
%\\[2mm]
%\PB{55mm}{$\sim 10^{-2}$} \hfill
%\PB{55mm}{$\sim 10^{-2}$}
%\\[2mm]
%\FB{\PB{55mm}{ $\omega\phi$ \\ $6.3(23)\cdot 10^{-4}$}} 
%\HS{5mm} VDM \HS{5mm} 
%\FB{\PB{55mm}{ $\rho\phi$   \\ $3.4(8)\cdot 10^{-4}$}}    % gas/P
%\\[5mm]
%\FB{\PB{55mm}{ $\gamma\phi$ \\ $1.7(4)\cdot 10^{-5}$}}
%\EC

%%%%%%%%%%%%%%%%%%%%%%%%%%%%%%%%%%%%%%%%%%%%%%%%%%%%%%%%%%%%%%%%%%%%
\section{Two-Step Mechanism as Dynamical Breaking of the OZI Rule}

Reactions which are OZI forbidden in a tree approximation can proceed 
via two-step mechanisms without a violation of
the OZI rule at the individual steps \cite{Li84,Li86,Li87,Li89}.    
Before discussing this mechanism in the $p\bar{p}$ annihilation
we briefly illuminate it by the example of the $\phi\to\rho\pi$ decay. 

%%%%%%%%%%%%%%%%%%%%%%%%%%%%%%%%%%%%%%%%%%%%%%%%%%%%%%%%%%%%%%%%%%%%
\subsection{The OZI rule violation $\phi\to\rho\pi$ decay}

  The width of the $\phi\to\rho\pi$ decay due to the $\omega-\phi$
mixing is given by the estimate  
\BA
   \Gamma_{\phi\to\omega\to\rho\pi} \approx 
   (\Theta-\Theta_i)^2
   \frac{g_{\omega\rho\pi}^2 P_{\phi\to\rho\pi}^3}{12\pi}
   \approx 0.16 \;\MB{\rm MeV}              
\EA
where $g_{\omega\rho\pi}^2/4\pi=20.6\;$GeV$^{-2}$ \cite{GMSW} and
$P_{\phi\to\rho\pi}$ is the CMS momentum of the particles in the final
state.  
This is significantly smaller than the experimental value
$\Gamma_{\phi\to\rho\pi}=0.57\;$MeV. 
The latter can be explained by the two step mechanism
$\phi\to K\bar{K}\to\rho\pi$ \cite{Pa,TW}.
% (Fig.~\ref{TSphi})  
%%%%%%%%%%%%%%
%\BFig
\VS{-20mm}
\BC \mbox{\epsfysize=50mm\epsffile{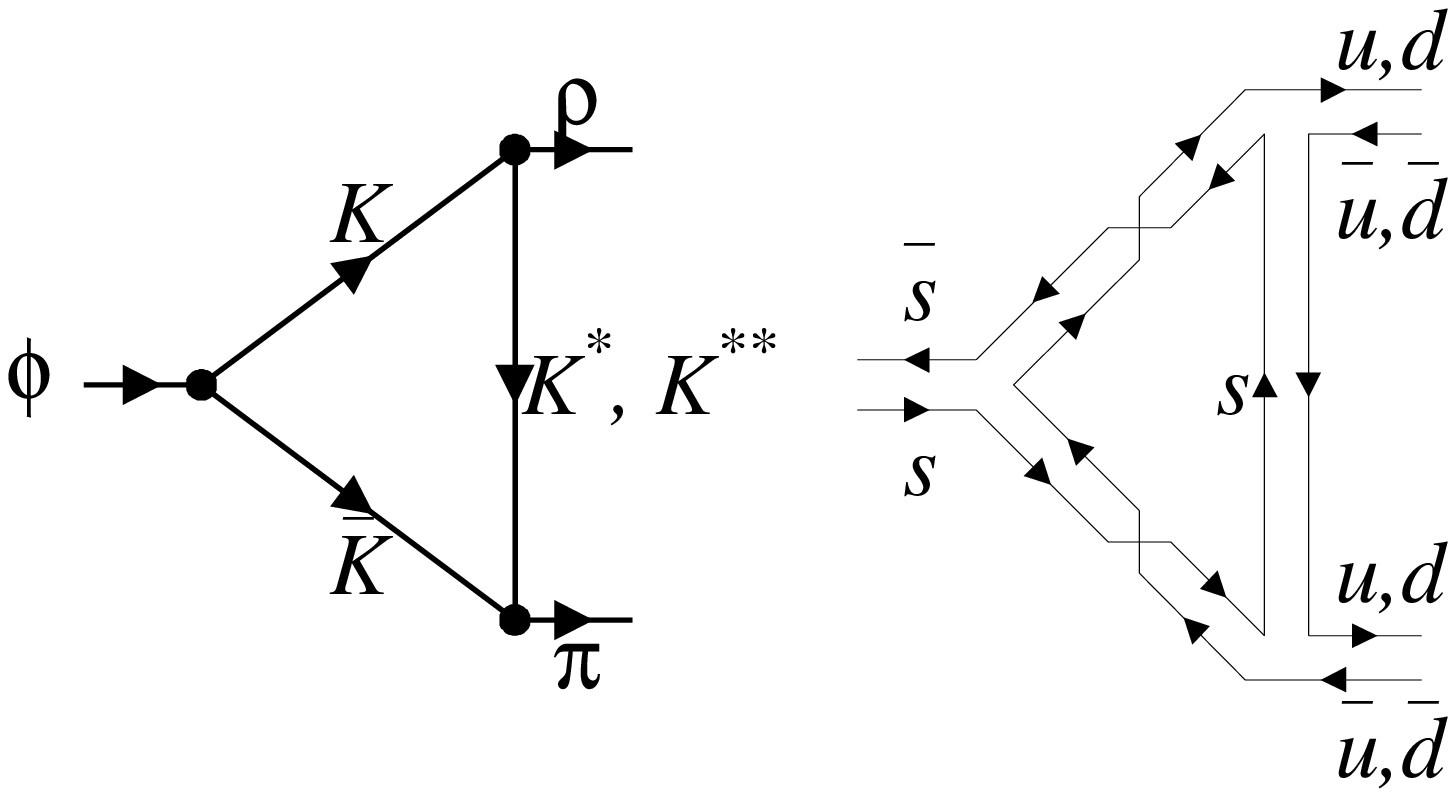}} \EC
\VS{-40mm}
%\caption{\LL{TSphi} The two step mechanism of the
%$\phi\to\rho\pi$ decay.}
%\VS{-5mm} 
%\EFig
%%%%%%%%%%%%%%
All the vertices are OZI allowed in this two step process. 
A simple estimate of the amplitude can be obtained in the unitarity
approximation when the intermediate particles $K\bar{K}$ are on the mass
shell. It gives the imaginary part of the amplitude 
which is of the right order of magnitude, but for a quantitative agreement
the real part is essential, and the full calculation of the hadronic
loops is needed \cite{Pa,TW}.

%%%%%%%%%%%%%%%%%%%%%%%%%%%%%%%%%%%%%%%%%%%%%%%%%%%%%%%%%%%%%%%%%%%
\subsection{Lipkin cancellations in hadronic loops}

Since two-step (one-loop) mechanisms tend to be large, one faces  the problem
of avoiding too large corrections to the OZI rule
\cite{Li84,Li86,Li87,Li89,GI,LZ96}. It
turns  out that cancellations between various intermediate states can
suppress the two step corrections in some cases. 

The effect of cancellations for the mass operator in the meson nonets is
demonstrated in Table~\ref{TabHLC} showing the relative sign of the various
kaonic intermediate states contributing to the real part of the mixing
amplitude in different $J^{PC}$ channels \cite{LZ96}.
Nearly perfect cancellation between the
$K\bar{K},K^*\bar{K},K\bar{K^*},K^*\bar{K^*}$ terms explains a small
deviation from the ideal mixing for $J^{PC}=1^{--},2^{++},3^{--}$ mesons,
while no cancellation occurs in the scalar sector where the mixing
is known to be large.    

%%%%%%%%%%%%%%%%%%%%%%%%%%%%%%%%%%%%%%%%%%%%%%%%%%%%%%%%%%%%%%%%%%%%%%%
\BTab[htb]
\caption{\LL{TabHLC} The relative sign of various intermediate states
in the hadronic loop contributing to the $q\bar{q} - s\bar{s}$ mixing
\cite{LZ96} and the corresponding deviation from the ideal mixing
$(\Theta -\Theta_i)$ (deg).}
\BC
\BT{|l|cccc|c|} \hline \VS{1mm}
 $J^{PC}$ & $K\bar{K}$ & $K\bar{K}^*$ & $K^*\bar{K}$ & $K^*\bar{K}^*$ &
 $\Theta -\Theta_i$  \\ \hline
 $1^{--}$  & + & -- & -- & + & 0.7 -- 3.4 \\
 $2^{++}$  & + & -- & -- & + &   7 -- 9 \\
 $3^{--}$  & + & -- & -- & + &   6 -- 7 \\ \hline
 $1^{+-}$  & 0 & -- & -- & + & $\sim 18$  \\
 $1^{++}$  & 0 & -- & -- & + & $\sim 26$  \\
 $0^{-+}$  & 0 & -- & -- & + &  45 -- 58  \\
 $0^{++}$  & + & 0  & 0  & + & $\sim 36$  \\ \hline
\ET
\EC
\ETab
%%%%%%%%%%%%%%%%%%%%%%%%%%%%%%%%%%%%%%%%%%%%%%%%%%%%%%%%%%%%%%%%%%%%%%%

A special case is the mass region above one and only one OZI allowed
channel threshold. In this case the corresponding loop has an imaginary part
which cannot be canceled by other terms. Therefore, with a sufficiently
strong coupling to a single OZI allowed channel, dynamical breaking of
the OZI rule can occur. The decay $\phi\to\rho\pi$ is a particular example
of this effect.  In the following section we consider how the two-step
mechanism works in the case of $p\bar{p}$ annihilation.

%%%%%%%%%%%%%%%%%%%%%%%%%%%%%%%%%%%%%%%%%%%%%%%%%%%%%%%%%%%%%%%%%%%%
\section{Two Meson Doorway Rescattering Mechanism
         in $p\bar{p}\to\phi\pi$}

  The OZI rule violation in the $p\bar{p}\to\phi\pi$ at rest is rather strong:
$BR(p\bar{p}\to\phi\pi) /BR(p\bar{p}\to\omega\pi) = % 0.096 \pm 0.015$
0.096 \pm 0.015$ \cite{CBC93b,CBC95,CBC95a}. 
  The two-step mechanisms in  $p\bar{p} \to \phi +X$ were studied in
\cite{LLZ,LL,BL94,BL94a,BL95,GLMR}. The most important intermediate
states are $K\bar{K^*}-K^*\bar{K}$ and $\rho\rho$, since they provide 
the maximum combined strength of the vertex functions
(Fig.\ref{TSpiphi}). 
In the $\rho\rho$ case the strong annihilation step \cite{CLZ} (a hundred
times bigger than the $\pi\phi$ branching ratio) is followed by the strong
$\rho\pi\pi$ vertex and the $\phi\rho\pi$ vertex of moderate strength. 
In the $K\bar{K}^*$ case the annihilation step is a factor of 10
weaker \cite{Co67}, but it is followed by maximum strength for the remaining
vertices in the second step.

%%%%%%%%%%%%%%%%%%%%%%%%%%%%%%%%%%%%%%%%%%%%%%%%%%%%%%%
\BFig[hb]
\VS{-10mm}
\BC \mbox{\epsfysize=50mm\epsffile{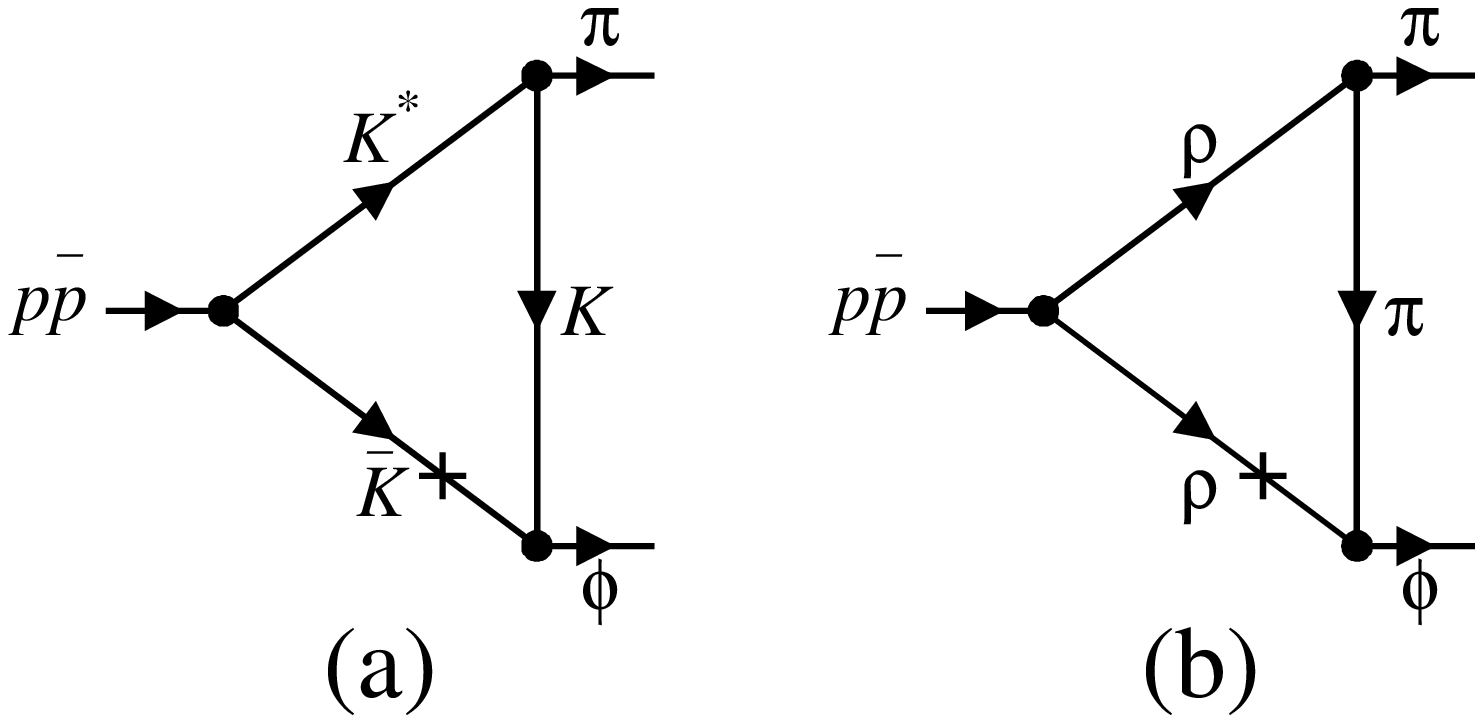}} \EC
\VS{-60mm}
\caption{Two-step mechanisms in $p\bar{p} \to \phi +X$.}
\LL{TSpiphi}
\EFig
%%%%%%%%%%%%%%%%%%%%%%%%%%%%%%%%%%%%%%%%%%%%%%%%%%%%%%%

 The scalar invariant amplitude has the form (see \cite{GLMR} for details) 
\BA
   G(s) = \frac{i}{\pi}  \int\limits_{0}^{+\infty}
   \frac{K dK}{E_1(K)} \, \frac{A(K,s) F_a}{(E_{1U}-E_1(K)+i\epsilon)} ,
   \label{GA}
\EA
where $A(K,s)$ is the spectral density shown in Fig.\ref{Apiphi} and
$F_a$ is the form-factor describing the combined off-mass-shell effects
for the intermediate state.  

%%%%%%%%%%%%%%%%%%%%%%%%%%%%%%%%%%%%%%%%%%%%%%%%%%%%%%%
\BFig[hbt]
\VS{-20mm}
\BC
\mbox{\epsfysize=70mm\epsffile{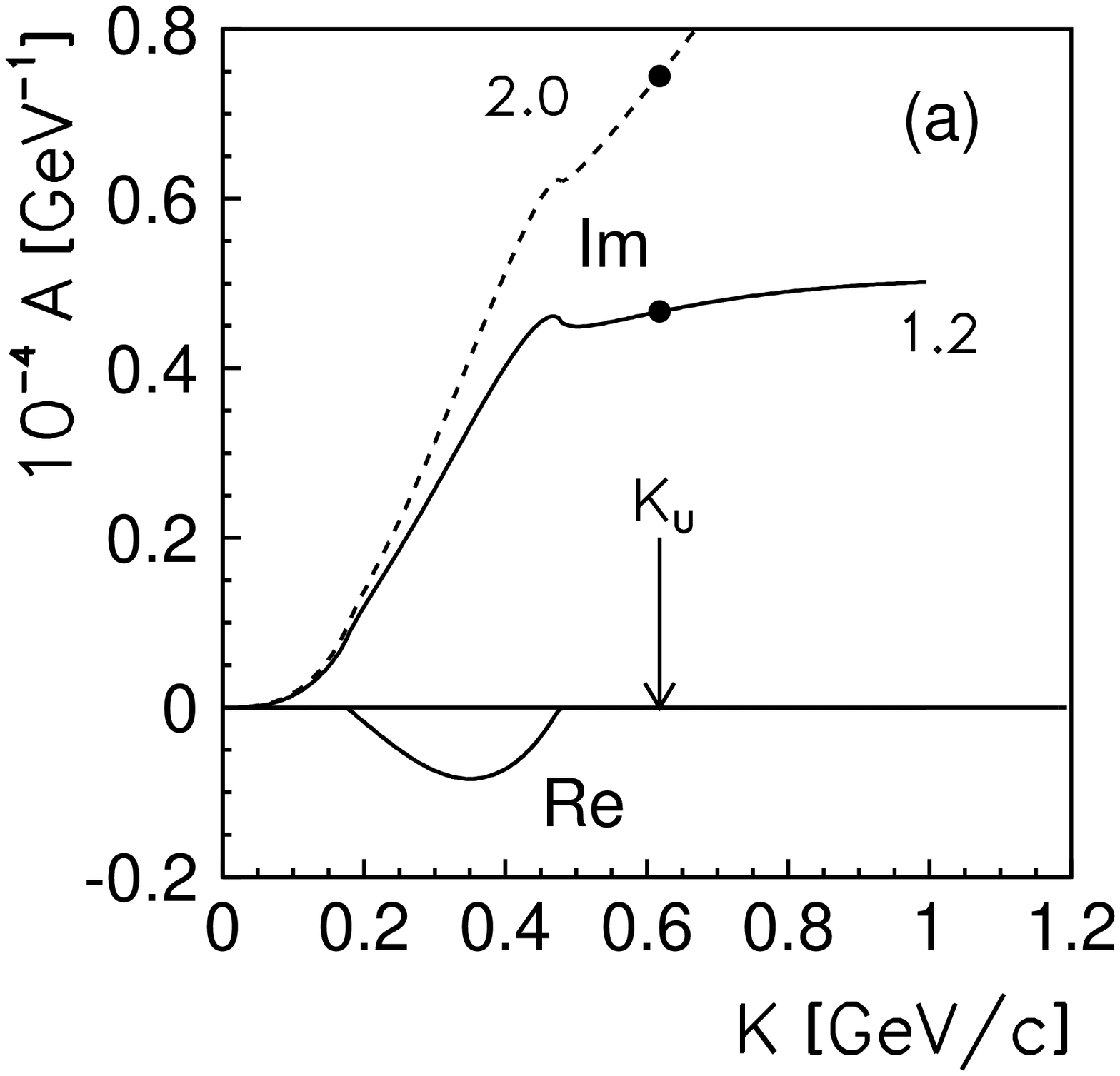}}
\HS{10mm}
\mbox{\epsfysize=70mm\epsffile{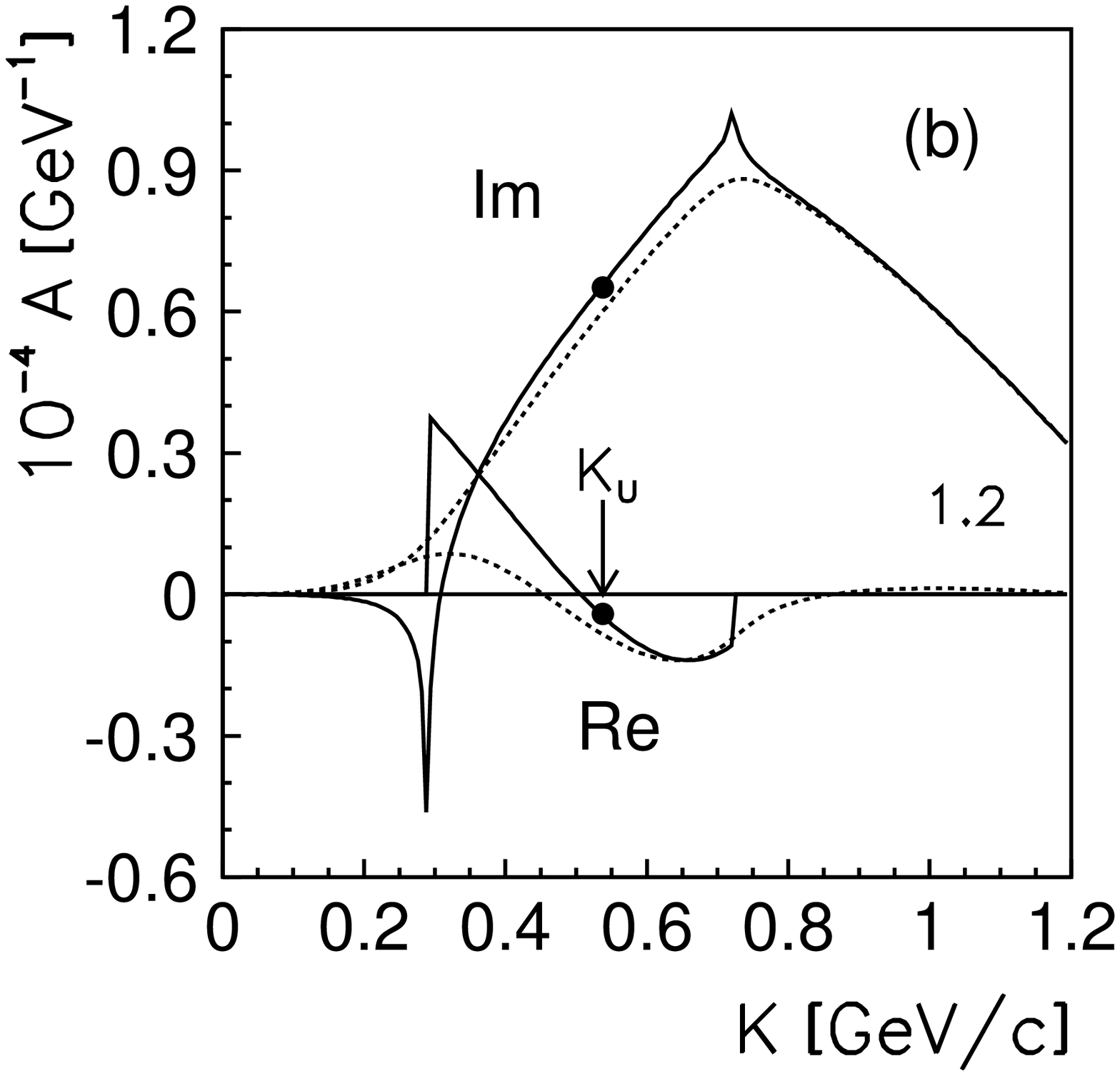}}
\EC
\VS{-40mm}
\caption{The spectral density $A(K,s)$ at threshold,
$s=4m_p^2$, {\it vs.} momentum $K$:
(a) the $K^*\bar{K}\,$-case, (b) the $\rho \rho\,$-case.  The unitarity
approximation amplitude $T_U=A(K_U,s)$ is marked by the full dots.
The two curves for the imaginary part plotted in (a) correspond to
different values of the form-factor parameter $\lambda_b$ (indicated in
units of GeV) describing the $t$-channel exchange in the second step
\protect\cite{GLMR}.
The dotted lines in (b) correspond to a Breit-Wigner propagator for the
$\rho$ meson.}
\LL{Apiphi}
\EFig
%%%%%%%%%%%%%%%%%%%%%%%%%%

The results of the one loop calculations are compared with the unitarity
approximation and the experimental data in Table~\ref{TabBRphipi}. The
relative sign of the $K\bar{K^*}$ and $\rho\rho$ terms is not known. In the
case of constructive interference the two step mechanism is in a good
agreement with the data.  Lipkin cancellations are not likely for this
reaction \cite{GLMR}.

%%%%%%%%%%%%%%%%%%%%%%%%%%%%%%%%%%%%%%%%%%%%%%%%%%%%%%%%%%%%%%%%%%%
\BTab[hbt]
\caption{\label{TabBRphipi} Branching ratio for
$p\bar{p}(^3S_1)\to\phi\pi^0$ in the unitarity approximation (UA) and
with the off-mass shell effects taken into account (1-loop) in comparison
with experiment \cite{CBC95}.  The form-factor parameters
$\lambda_a$ and $\lambda_b$ \protect\cite{GLMR} are in GeV.}
\BC
\BT{|l|l|c|c|c|c|}
\HL
\multicolumn{2}{|c|}{Theory} & \multicolumn{4}{|c|}{Two step mechanism} \\
\cline{3-6}
\multicolumn{2}{|c|}{\ \ \ } &
\makebox[2cm]{$K^*\bar{K}$} &
\makebox[2cm]{$\rho\rho$} &
\makebox[2cm]{incoherent} &
\makebox[2cm]{maximum} \\
\hline
UA             & $\lambda_b = \infty $ &
                 $3.0\times 10^{-4}$ & $2.0\times 10^{-4}$
               & $5.0\times 10^{-4}$ & $9.9\times 10^{-4}$ \\
               & $\lambda_b = 1.2 $    &
                 $0.63\times 10^{-4}$ & $0.33\times 10^{-4}$
               & $0.96\times 10^{-4}$ & $1.87\times 10^{-4}$ \\
\hline
1-loop         & $\lambda_a = 1.5$,  $\lambda_b = 1.2 $    &
                 $0.55\times 10^{-4}$ & $0.53\times 10^{-4}$
               & $1.09\times 10^{-4}$ & $2.16\times 10^{-4}$ \\
               & $\lambda_a = 2.0$, $\lambda_b = 1.2 $    &
                 $0.69\times 10^{-4}$ & $0.53\times 10^{-4}$
               & $1.22\times 10^{-4}$ & $2.40\times 10^{-4}$ \\
               & $\lambda_a = 2.0$,  $\lambda_b = 1.5 $ &
                 $1.59\times 10^{-4}$ & $0.85\times 10^{-4}$
               & $2.44\times 10^{-4}$ & $4.60\times 10^{-4}$ \\
\hline\hline
 \multicolumn{2}{|c|}{Crystal Barrel Collab.} 
 & \multicolumn{4}{|c|}{$(5.5 \pm 0.7)\times 10^{-4}$} \\
%%% & \multicolumn{4}{|c|}{$(4.5 \pm 1.1)\times 10^{-4}$} 
\HL
\ET
\EC % \VS{10mm}
\ETab
%%%%%%%%%%%%%%%%%%%%%%%%%%%%%%%%%%%%%%%%%%%%%%%%%%%%%%%%%%%%%%%%%%%

%The question about the energy dependence of the two-step mechanism
%remains open because of large theoretical uncertainties for the
%higher partial waves. 

Apart from annihilation at threshold the energy dependence carries
potentially important information on the OZI rule violation. The energy
dependence of the two step mechanism for $p\bar{p}\to\phi\pi$ has been
estimated in \cite{BL94a}. The predictive power in this case suffers from
too many unknown parameters.

%%%%%%%%%%%%%%%%%%%%%%%%%%%%%%%%%%%%%%%%%%%%%%%%%%%%%%%%%%%%%%%%%%%%
\section{Two Meson Rescattering Mechanism in $p\bar{p}\to\phi\phi$}

  This reaction was studied in \cite{LZL} in the unitarity approximation.
The most important intermediate state is $K\bar{K}$ (Fig.\ref{TSphiphi}a)
for it has the maximum coupling to the final state.  
Notice that the $\phi\phi$ production via the $\omega-\phi$ mixing
is strongly suppressed: 
$\sigma_{p\bar{p}\to\omega\omega\to\phi\phi} 
/\sigma_{p\bar{p}\to\omega\omega} \sim (\Theta-\Theta_i)^4 \sim 10^{-5}$. 
The calculations of the two-step mechanism have correct magnitude in
comparison with the experiment \cite{JETSET} as shown in Fig.\ref{TSphiphi}b.
More data concerning the energy dependence of the $\phi\phi$ production
would be very helpful for a further test of the mechanism considered.     

%%%%%%%%%%%%%%%%%%%%%%%
\BFig[h]
\BC (a) \HS{100mm} (b) \EC
\VS{-25mm}
\BC
\mbox{\epsfysize=25mm\epsffile{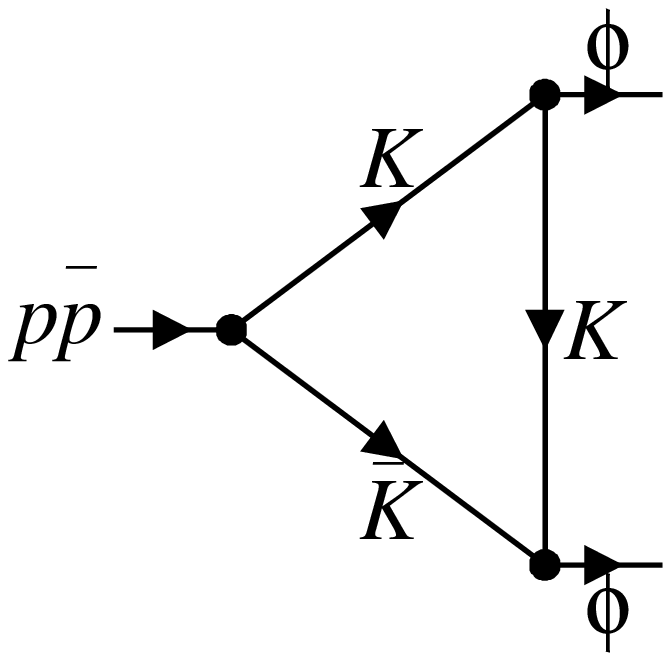}} \HS{10mm}
\mbox{\epsfysize=25mm\epsffile{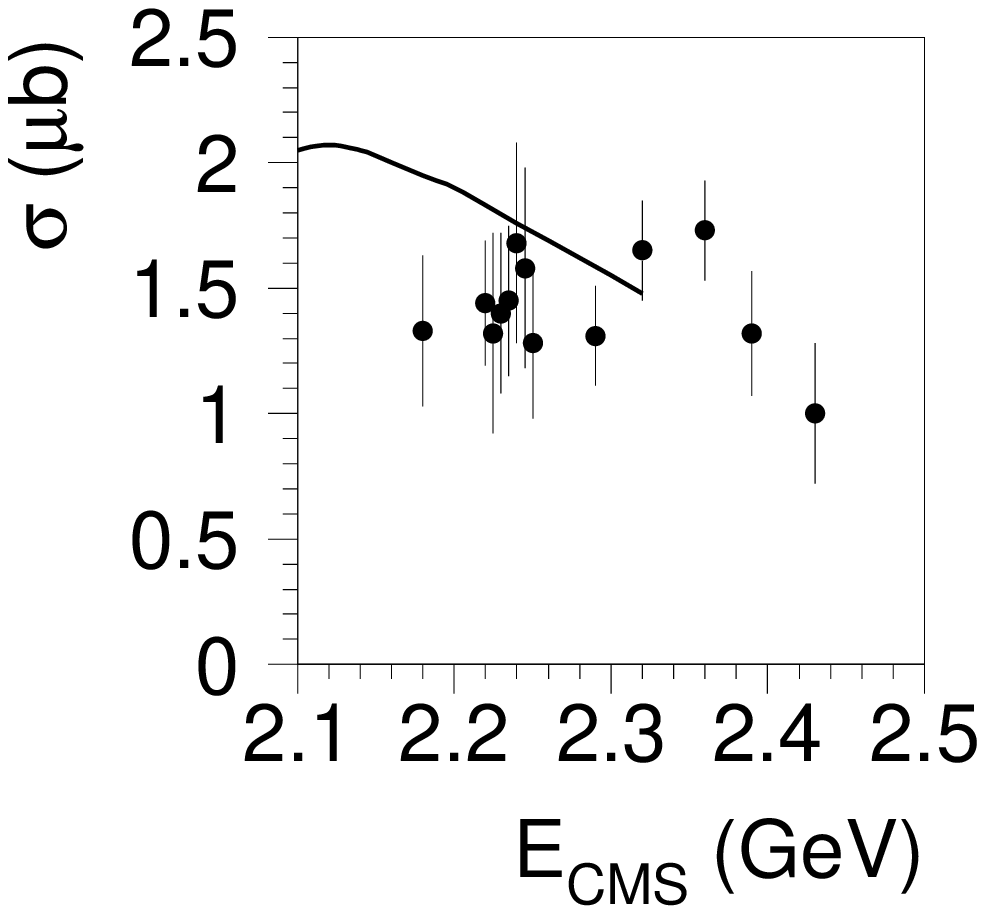}} 
\EC
\VS{-40mm}
\caption{(a) The $K\bar{K}$ rescattering mechanism in $p\bar{p}\to\phi\phi$.
(b) The energy dependence of the $p\bar{p}\to\phi\phi$ cross section
calculated for the two step mechanism \protect\cite{LZL}.
Experimental data from \protect\cite{JETSET}.}
\LL{TSphiphi}
\EFig
%%%%%%%%%%%%%%%%%%%%%%%

%%%%%%%%%%%%%%%%%%%%%%%%%%%%%%%%%%%%%%%%%%%%%%%%%%%%%%%%%%%%%%%%%%%%
\section{The OZI Violation in
         $p\bar{p}\to\phi\pi^+\pi^-,\phi\rho,\phi\omega$}
\LL{TSrhophi}

Table~\ref{OZIphirho} summarizes the experimental data on the OZI rule
violation in the $\phi\pi^+\pi^-$ and $\phi\rho$ channels at different
energies. The OZI rule violation in these channels is not dramatic compared
with the level given by the $\omega-\phi$ mixing
$(\Theta-\Theta_i)^2\sim 4\cdot 10^{-3}$.
The measured branching ratios for the reactions $p\bar{p}\to\phi\rho,
\phi\omega$ at rest are given in Tab.~\ref{TabBRVV}. 
The OZI rule violation in the $\phi\omega$ channel,  
$BR(\phi\omega)/BR(\omega\omega)=(19\pm 7)\cdot 10^{-3}$ (liquid) 
\cite{CBC93b,Biz71}, is comparable with the effect in the $\phi\rho$
channel. 

%%%%%%%%%%%%%%%%%%%
\BTab[hbt]
\caption{\label{OZIphirho} The relative probabilities for
the $\phi\rho,\omega\rho$ and $\phi\pi^+\pi^-,\omega\pi^+\pi^-$
channels in the $p\bar{p}$ annihilation {\it vs.} antiproton momentum
(in units $10^{-3}$).} 
\BC
\BT{|l|c|c|l|}
\HL
$p_{\bar{p}}$(GeV/c) & $\frac{BR(\phi\rho)}{BR(\omega\rho)}$ 
& $\frac{BR(\phi\pi^+\pi^-)}{BR(\omega\pi^+\pi^-)}$ & Ref. \\
\HL
 0 (gas)    & $6.3 \pm 1.6$ &                    & \cite{AST91,AST93} \\
 0 (gas/LX) & $7.5 \pm 2.4$ &                    & \cite{AST91,AST93} \\
 0 (liq.)   &               & $7.0 \pm 1.4$      & \cite{Biz69} \\
 0.76       & $9 \pm 5$     & $10.0\pm 2.4$      & \cite{Co78} \\
 1.2        &               & $11 {+3 \atop -4}$ & \cite{Do76} \\
 2.3        & $22 \pm 13$   & $21 \pm 5$         & \cite{Ch77} \\
 3.6        &               & $9 {+4\atop -7}$   & \cite{Do76} \\  
%%%% phase space corrected:
% 0 (liq.)   &               & $15 \pm 3$         & \cite{Biz69} ?? \\
% 0.76       & $13 \pm 4$    & $19 \pm 5$         & \cite{Co78} \\
% 1.2        &               & $19 {+5 \atop -7}$ & \cite{Do76} \\
% 2.3        & $25 \pm 15$   & $30 \pm 7$         & \cite{Ch77} \\
% 3.6        &               & $12 {+5\atop -9}$  & \cite{Do76} \\  
%%%%%
\HL
\ET
\EC
\ETab
%%%%%%%%%%%%%%%%%%%%

The two-step mechanisms with two particles in the intermediate state were
considered in the unitarity approximation in \cite{BL94} where the
calculated branching ratios for the $\phi\rho$ and $\phi\omega$ channels
were found to be almost two orders of magnitude smaller than the data. The
off-mass-shell corrections have not yet been calculated for these processes.
It is also not excluded that the intermediate states with more than two 
particles (like $\rho\rho\pi, K^*\bar{K}\pi$ in the case of $\phi\rho$)
are important.   

%%%%%%%%%%%%%%%%%%%%%%%%%%%%%%%%%%%%%%%%%%%%%%%%%%%%%%%%%%%%%%%%%%%%%%%%%
\section{Conclusion}

Two step mechanisms play an important role in the observed violations
of the OZI rule in the $p\bar{p}$ annihilation at low energies.  
\\ 
-- The two meson rescattering mechanisms for $\bar{p}p \to \pi\phi$
including all off-mass-shell effects have a size almost two orders of
magnitude bigger than the OZI tree level expectations, in agreement
with the measured branching ratios at rest.
The $K^*\bar{K}$ rescattering mechanism does not request any unexplained OZI
violation for the  individual vertices in the diagrams. 
\\
-- The $\phi\phi$ production in the $p\bar{p}$ annihilation at
$\sqrt{s}<2.5\;$GeV can be explained by the two step mechanism with the
intermediate state $K\bar{K}$.
\\
-- The $\phi\gamma$/$\omega\gamma$ ratio in the $p\bar{p}$ annihilation at rest
can be explained by the VDM. The $(\rho\!\to\!\gamma)+(\omega\!\to\!\gamma)$
interference is destructive for $\omega\gamma$ and constructive for
$\phi\gamma$.

Thus the most drastic violations of the OZI rule can be explained by the
known mechanism of rescattering. In particular no substantial $s\bar{s}$
component in the nucleon is required. Further studies of the
OZI rule breaking due to two step corrections in other
annihilation channels ($\phi\rho$, $\phi\pi\pi$,$\phi\omega$) and the energy
dependence of the $\phi$ production would be very desirable.  

%%%%%%%%%%%%%%%%%%%%%%%%%%%%%%%%%%%%%%%%%%%%%%%%%%%%%%%%%%%%%%%%%%%%%%%%%
\section*{Acknowledgements}
The author would like to thank M.P.~Locher and S.~von~Rotz for
fruitful collaboration and O.~Gorchakov, F.~Lev, and M.~Sapozhnikov
for useful discussions. 

%%%%%%%%%%%%%%%%%%%%%%%%%%%%%%%%%%%%%%%%%%%%%%%%%%%%%%%%%%%%%%%%%%%%%%%%%

%%%%%%%%%%%%%%%%%%%%%%%%%%%%%%%%%%%%%%%%%%%%%%%%%%%%%%%%%%%%%%%%%%%%%%%%%

\end{document}